\documentclass[twocolumn]{aastex63}
\usepackage{multirow}
\usepackage[T1]{fontenc}
\usepackage{mhchem}
\usepackage{url}
\usepackage{float}
\usepackage{bbding}
\usepackage{pifont}
\usepackage{wasysym}
\usepackage{amssymb}

\floatstyle{plaintop}

\usepackage{graphics,graphicx}

\newcommand{\HtwoO}{\(\textrm{H}_{2}\textrm{O}\)}
\newcommand{\Msun}{\(\textrm{M}_{\odot}\)}

\newcommand{\midUV}{\(10^{4} \textrm{G}_{0}\)}
\newcommand{\highUV}{\(10^{6} \textrm{G}_{0}\)}

\newcommand{\CtwoHtwo}{\textrm{C}$_{2}$\textrm{H}$_{2}$}

\definecolor{LightYellow}{rgb}{1,.98,.804}

\submitjournal{APJ}

\begin{document}
\title{The Impact of External Radiation on the Inner Disk Chemistry of Planet Formation}

\shorttitle{UV Backgrounds}


\author[0000-0002-0150-0125]{Jenny K. Calahan}
\affiliation{Center for Astrophysics | Harvard \& Smithsonian, 60 Garden St., Cambridge, MA 02138, USA }

\author[0000-0002-0150-0125]{Karin \"{O}berg}
\affiliation{Center for Astrophysics | Harvard \& Smithsonian, 60 Garden St., Cambridge, MA 02138, USA }

\author[0000-0002-0150-0125]{Alice Booth}
\affiliation{Center for Astrophysics | Harvard \& Smithsonian, 60 Garden St., Cambridge, MA 02138, USA }

\begin{abstract}

The vast majority of young stars hosting planet-forming disks exist within clustered environments, like the Orion Nebula, implying that seemingly `extreme' UV environments (10$^{4}$ G$_{0}$ and above) are not so atypical in the context of planet formation. Using thermo-chemical modeling, we explore how the temperature and chemistry within a protoplanetary disk around a T Tauri star is impacted by the surrounding UV environment. The disk becomes hotter due to heating by photodissociation of molecules, photoelectric heating, H$_{2}$, and atomic processes and as a result the area in which molecules exist in the ice-phase shrinks, being pushed both downward and inward. Beyond 1~AU the chemistry changes most significantly in a UV-rich background; the atmosphere becomes more H$_{2}$O, OH, and atomic-rich. Hydrocarbons, however, reside primarily well within 1~AU of the disk, thus their abundance and distribution is not impacted by the UV field, up to a \highUV{}. 
The products of photodissociation and photochemistry are formed deeper into the disk with increasing UV background field strength beyond 1~AU, impacting the chemistry near the midplane. Effectively a `reset' chemistry takes place, with an enhancement of atoms, simple molecules, and molecules in the gas-phase. Planets that form in highly irradiated regions will be exposed to a different chemical reservoir in the gas and ice-phases than that in an isolated disk, and the impact from the UV background should only be detectable in highly irradiated disks ($\sim$\highUV{}). 


\end{abstract}


\section{Introduction} \label{sec:intro}

Protoplanetary disks harbor the building blocks to form planets; understanding the underlying chemistry provides connections between star and planet formation. Most protoplanetary disks that we have studied in depth are isolated sources \citep[i.e.][]{Andrews18, Oberg21}, as they are nearby and easier to observe. However, our solar system formed in a clustered environment \citep{Adams10}, much like the vast majority of stars and thus planets \citep{Lada03, Parker20}.  
One of the strongest environmental differences between a disk within a clustered environment, and one in isolation, is the background radiation field. In a densely populated star-forming region, low-mass to solar-mass stars will form alongside O- and B-stars, which greatly enhance the local UV radiation field impinging on the disks surrounding lower-mass stars. The external UV background can impact a disk's chemistry, thermal structure, external physical structure (truncated disk, ionized jets, winds, envelopes), and dynamics \citep[i.e.]{Walsh13, Keyte25, UVIrr25}. 

Planet formation in such clustered environments has become increasingly accessible observationally, mainly due to a combination of JWST and ALMA programs. From the perspective of ALMA, on a global scale, the Orion nebula's star forming efficiency and basic dust disk parameters are on par with other nearby star forming regions \citep{Ansdell20, vanTerwisga22}. Focusing on disks near a massive O-star within the Orion nebula, it has been seen that both the dust-disk and gas-disk are more compact than those in more isolated environments \citep{Eisner18, Boyden20, vanTerwisga20, vanTerwisga23}.
There appears to be a slight correlation between disk size and distance from the bright O-star, suggesting that disks are being truncated via photo-evaporation due to the environment \citep{Haworth17}. 
Further evidence for this theory comes from the fact that highly irradiated disks appear to be less massive than their non-irradiated counter parts \citep{Ansdell17, Mauco23} likely via mass-loss via photoevaporative winds \citep{ConchaRamirez21}.  These features of living within a $>$10$^{3}$G$_{0}$ environment \citep[1 G$_{0}$ = 1.6$\times 10^{-3}$ ergs cm$^{-2}$ s$^{-1}$][]{Habing68} will certainly impact the potential for planet formation in the outer disk directly. 

The impact on the chemistry of planet formation from a high UV background has not been as thoroughly explored. With ALMA, two disks in the outskirts of the Orion nebula appear to have similar chemical signatures as isolated disks \citep{Diaz-Berrios24}. Models of moderately irradiated disks suggest that from the outside-in, the disk temperature structure will increase both in the midplane ($\sim$2x) and atmosphere (orders of magnitude hotter), and the bulk of the chemical changes will occur at 100~AU as compared to 10~AU\citep{Walsh13, Boyden23, Gross25}. Studying the chemical effects observationally at mm-wavelengths on $<10$~AU scales is difficult, as they are located far away. The inner few AU of the disk is more accessible using IR telescopes, and a small sample of disks within irradiated environments have a published JWST spectra.


The inner few AU contains the highest density gas and dust, harbors the water snowline, and it coincides with the majority of terrestrial planet formation, and likely super-Earth to Neptune-like planets. The impact on the chemistry of planet formation within the inner few AU from a UV irradiated environments remains unclear. A proplyd (d203-506) and a envelope free disk (XUE 1) are the first two irradiated sources with published data, with distinctly different signatures. The proplyd is found in a $4 \times$10$^{4}$G$_{0}$ environment, and has a detection of CH$_{3}^{+}$ \citep{Berne23} which is indicative of active UV chemistry in the atmosphere and a solar C/O ratio within the inner AU \citep{Schroetter25}. XUE 1 is predicted to have been exposed to 10$^{5}$ G$_{0}$, and looks remarkably similar to non-irradiated disks, and it exhibits all common molecular signatures: H$_{2}$O, CO, CO$_{2}$, HCN, C$_{2}$H$_{2}$ \citep{Ramirez-Tannus23}. XUE 1 does, however, appear to be enhanced in water emission beyond 10~AU as compared to isolated disks \citep{Portilla-Revelo25}. Eleven more sources within the XUE sample have been observed and published \citep{RamireTannus25} with their spectra overall looking remarkably similar to non-irradiated sources. There are many observations of irradiated disks that have or will soon be completed, and a large sample is required to disentangle the impact due to the background UV from the physical structure, stellar type, and other source properties. Yet, we are lacking a interpretive framework from a thermo-chemical modeling perspective. As attention is turned towards JWST and understanding the chemistry of planet formation, it is critical that we explore the impact of the radiation background, focused on the inner few AU. 

In this study, we explore how external irradiation will directly impact the chemistry of the key planet-forming zone. We do this via a thermo-chemical model representing a `typical' disk around a K-star. We explore how the impact of external irradiation depends on the disk mass, and C/O ratio, and step through a radiative background equivalent to an isolated disk, up to a disk that resides within 0.05 pc of an O star (10$^{6}$ G$_{0}$). This upper limit represents a maximum UV irradiation background that is still typical of clustered environments \citep[i.e. Westerlund 1, Orion, Trumpler 14;][]{Winter18,Winter22}. In Section \ref{sec:method} we lay out the structure of the thermo-chemical code we use as well as the additional accretion module that is unique to this study. In Section \ref{sec:results} we present the main results as to how the UV background impacts temperature, the potentially-observable chemistry, and chemistry directly available to forming planets within 10~AU. In Section \ref{sec:diss} we explore the impact of these results on planet formation processes and provide an explanation behind the trends seen in Section \ref{sec:results}, and compare our results to those of current observations. And finally we conclude in Section \ref{sec:conclusion} with our take-away message and outlook on future modeling avenues. 

\section{Methods}\label{sec:method}

\subsection{Model Overview}
This work uses a 2D thermo-chemical model, \textsc{DALI} \citep{Bruderer12,Bruderer13}. \textsc{DALI} assumes a 2D physical structure including gas and dust densities, a stellar spectrum, and background radiation field, a chemical network, and initial abundances amongst other assumptions. The model then calculates a dust temperature and radiation field through out the disk. Within the inner disk, viscous heating from accretion likely is becoming a more significant heating source, impacting the location of molecules in the gas-phase, thus the chemistry available to planet formation. We add an accretion heating module to DALI, which is detailed in Section \ref{sec:Acc}. Once the dust temperature and radiation field are finalized, DALI then determines the gas temperature and chemical abundances iteratively. 2D gas and dust temperature and chemical abundances serve as final results.

\begin{deluxetable}{cccc}
\label{tab:Model_properties}
\tablecolumns{7}
\tablewidth{0pt}
\tabletypesize{\small}
\tablecaption{Model Variables}
\tablehead{Model Name	& Mass$^{\rm{a}}$ & C/H & O/H }
 \startdata
 Fiducial & 0.06 \Msun{} & 1.35$\times$10$^{-4}$ & 2.88$\times$10$^{-4}$\\
 Low Mass & 0.006 \Msun{} & 1.35$\times$10$^{-4}$ & 2.88$\times$10$^{-4}$\\
 High C/O & 0.06 \Msun{} & 1.35$\times$10$^{-4}$ &  6.76$\times$10$^{-5}$\\
 \enddata
\tablecomments{$^{\rm{a}}$ The mass listed is what would be contained within a disk with r$_{\rm{in}}$= 0.03~AU and r$_{\rm{out}}$= 300~AU, however we truncate our disks at 30~AU.\newline
Each model also is exposed to three UV fields: 1 G$_{0}$, \midUV{}, and \highUV{}}
\end{deluxetable}

The UV field throughout the disk is driven by the stellar and background UV. DALI initially assumes an  isotropic UV background that is applied evenly throughout the disk, following \citet{Draine78}. The dust and molecular column densities from the perspective of the star and the vertical direction are then used to determine how the external UV field impinges into the disk. To represent an increase in the radiation field due to a close distance to an O-star, we multiply this Draine radiation field by a factor of 10-10$^{6}$ throughout the wavelength range. We focus on results at 1 G$_{0}$, \midUV{}, and \highUV{}. We set out to create a fiducial model that would represent a `typical' disk surrounding a K-star in an irradiated environment. This disk has a total gas mass of 1\% of the stellar mass, a gas-to-dust ratio of 1000, and 99\% of the dust mass is in large grains (see Appendix for more information regarding the physical structure of the disk). 
This deviation from the typical gas-to-dust of 100, and 90\% of the mass in large grains is motivated by previous DALI models that reproduce Spitzer and JWST observations \citep{Bosman22a}. The chemical network is expanded from the default network typically used by DALI \citep[based on][]{Doty04, Stauber05,Bruderer09}, as some IR-emissive molecules are not included, most critically C$_{2}$H$_{2}$. The expanded network utilizes some of the molecules and their associated binding energies from \citep{Bosman18}. Not all could be included, otherwise the thermal and chemical balance calculation within the current version of DALI would not converge in a reasonable amount of time. We focused on expanding the hydrocarbon network, and only included molecules that had reactions in the UMIST06 database \citep{Woodall07}. The network was comprised with the goal of ensuring that C$_{2}$H$_{2}$ does not become a carbon sink, thus we include all hydrocarbons up to C$_{6}$H$_{6}$. We distribute our elements into atomic constituents to initialize the chemistry, and the PAH abundance is 6.0 $\times$10$^{-7}$/H, which is approximately the ISM abundance. The PAH abundance in protoplanetary disks is not a well constrained quantity, and can contribute to the heating of the disk. We find that by decreasing the PAH abundance by an order of magnitude results in a warmer disk atmosphere, by up to a factor of three.

To explore how the initial chemical and physical structure may impact our results, we also set up a carbon-rich disk model (initialized C/O = 2 instead of 0.47, per the fiducial model setup) and low-mass disk model (gas mass is 0.1\% of the stellar mass), see Table \ref{tab:Model_properties} for details. Both the carbon-rich case and lower mass disk case are motivated by observations of protoplanetary disks, where a wide range of C/O ratios have been seen both in the outer and inner disk \citep{Miotello19, Bosman21_c2h, Tabone23, Colmenares24} and in UV irradiated regions in particular, disks are found to be typically lower in mass\citep{vanTerwisga20}. 

\subsection{Accretion Heating Module}\label{sec:Acc}
Accretion of disk material on to the central star will provide an extra heating term to the inner disk that can be safely ignored beyond a few AU. However, this study focuses on inner disk chemistry, and it was critical to add accretion heating to the thermo-chemical modeling framework. This represents heating from midplane-based accretion, and its heat transfer upwards. The gas and dust temperature is a combination of the irradiation temperature and accretion temperature via

\begin{equation}
    T = (T_{\rm{irr}}^{4} + T_{\rm{acc}}^{4})^{1/4} .
\end{equation}

\noindent the irradiation temperature T$_{\rm{irr}}$ is calculated within DALI, and we allow T$_{\rm{acc}}$ to be calculated based on a given accretion rate and the vertical surface density. We use

\begin{equation}
    T_{\rm{acc}}^{4}=\frac{3}{8\pi}\dot{M}\Omega^{2}\frac{\tau_{\rm{vert}}}{\sigma},
\end{equation}

\noindent from \citet{DAlessio98, Lodato08, Harsono15} and references therein. Here, $\dot{M}$ is the assumed accretion rate constant across the midplane \citep[10$^{-8}~$M$_{\odot}$/yr, a typical accretion rate for Class II protostars,][]{Manara23}, $\Omega$ is the angular velocity, $\sigma$ is the stefan-boltzmann constant, and $\tau$ is the optical depth, defined in this context as 

\begin{equation}
    \tau_{\rm{vert}} = \frac{1}{2}\Sigma \kappa.
\end{equation}

\begin{figure}
    \centering
    \includegraphics[width=0.9\linewidth]{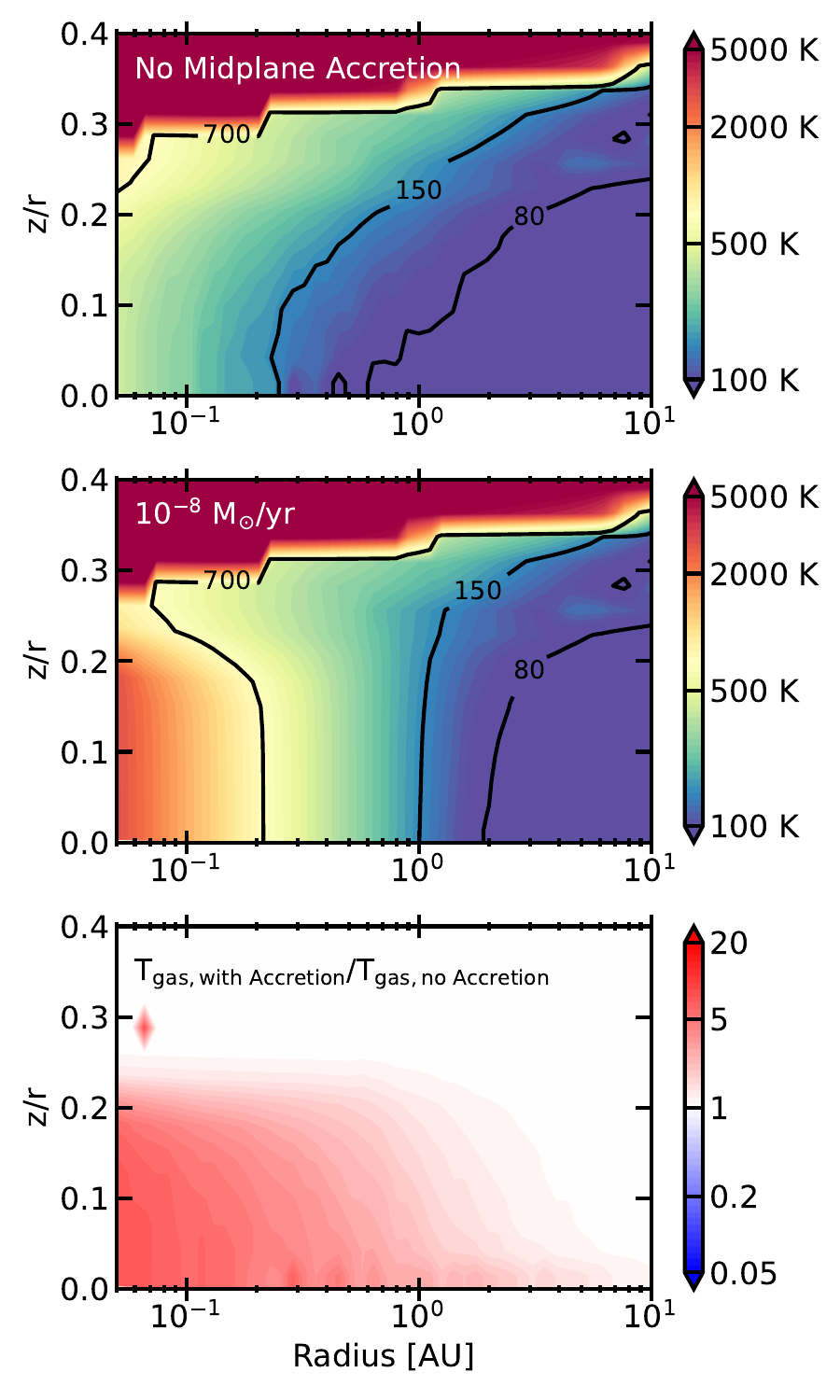}
    \caption{\textit{The temperature structure of a protoplanetary disk without accretion heating in the midplane [left], and with a 10$^{-8}$M$_{\odot}$/yr accretion in the midplane [right].}}
    \label{fig:accretion_example}
\end{figure}

\noindent $\Sigma$ is the surface density of the disk at a given radius, while $\kappa$ is an opacity value, which we use a constant value of 0.5 cm$^{2}$/g \citep{Miyake93}. We add this effective accretion heat source after the dust radiative transfer calculations, i.e. after T$_{\rm{irr}}$ is determined. The final dust and gas temperature at this step is then determined with the additional contribution from T$_{\rm{acc}}$. We find that in this implementation, the addition of accretion heating primarily impacts the midplane temperature, and any impact is kept within the warm molecular-rich zone (in this model, $\sim$z/r$<$0.2, see Fig.~\ref{fig:accretion_example}). The temperature near the midplane, closest to the star increases by a factor of 10, with decreasing impact farther from the star. This does result in a temperature inversion near the atmosphere (between z/r = 0.2-0.3 in the fiducial model) within 1~AU, where the midplane temperature is warmer than the upper extent of the molecular region. 

\section{Results}\label{sec:results}

\begin{figure*}
    \centering
    \includegraphics[width=0.7\linewidth]{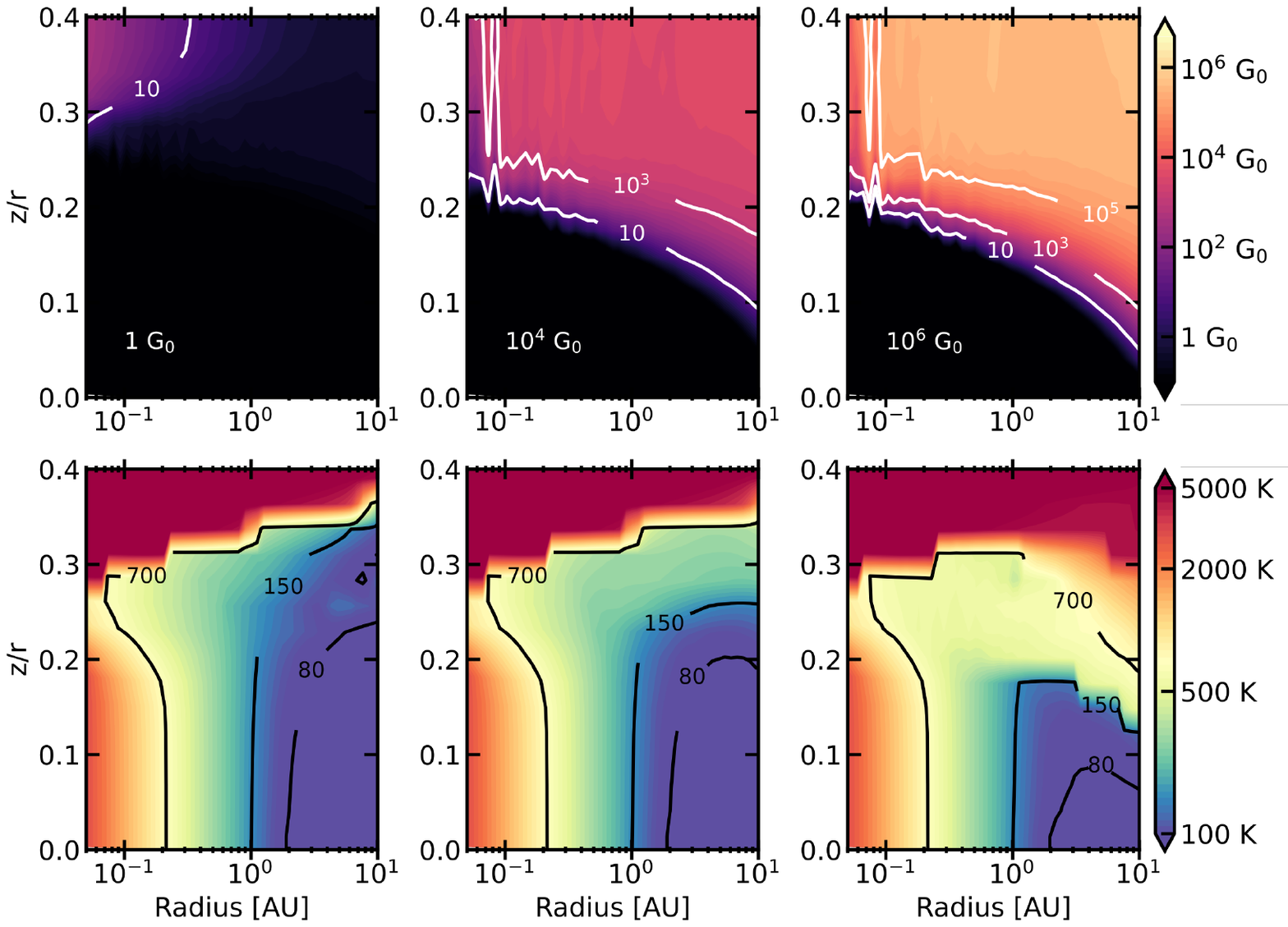}
    \caption{\textit{The internal UV field [top row] and temperature [bottom row] within the first 10 AU of a disk that is 1\% of the stellar mass (0.06 \Msun), exposed to an 1 G${_0}$, 10$^{4}$ G${_0}$, and 10$^{6}$ G${_0}$ background}}
    \label{fig:fiducial_tgas_evol}
\end{figure*}

\begin{figure*}
    \centering
    \includegraphics[width=0.7\linewidth]{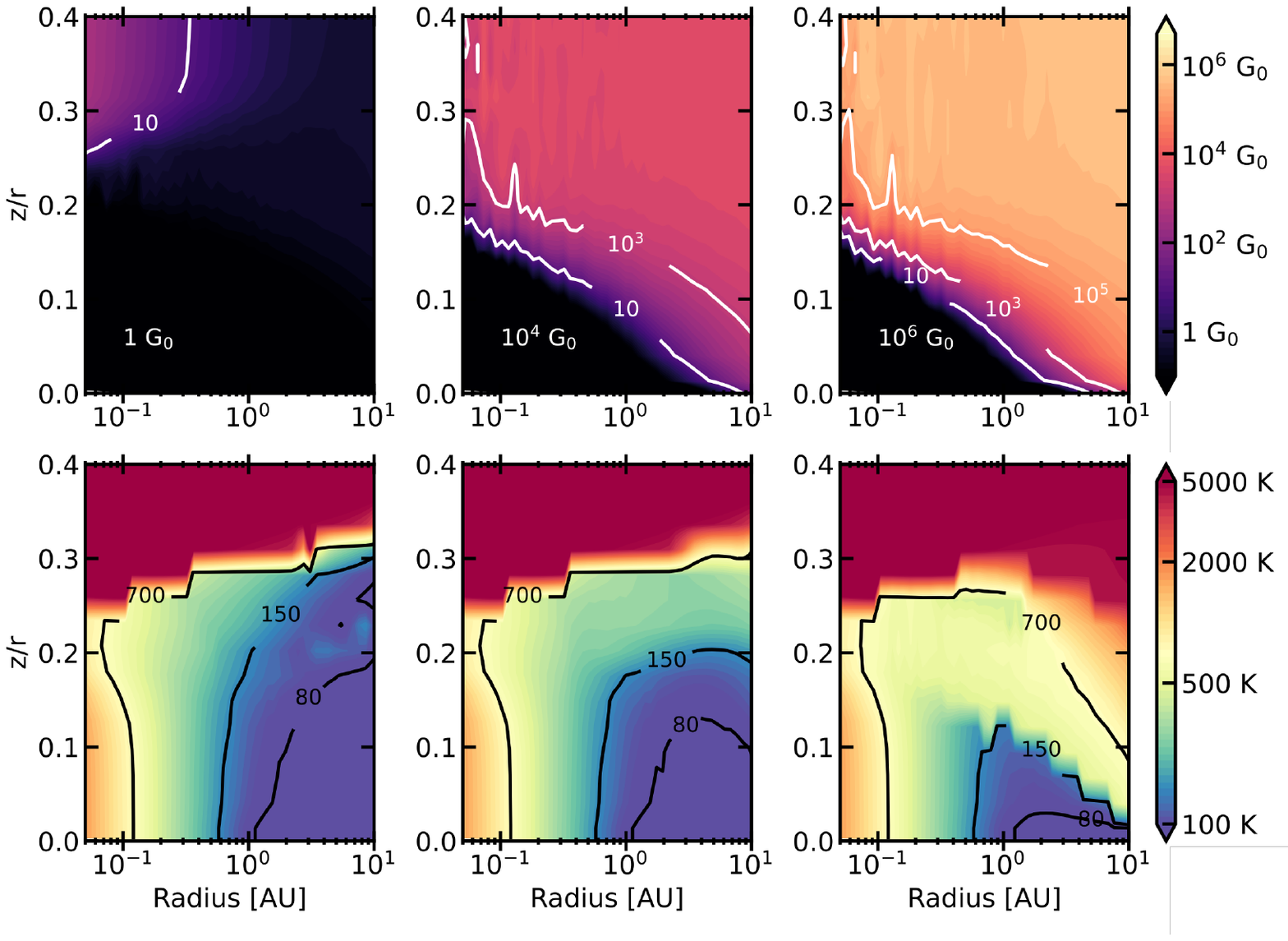}
    \caption{\textit{Same as Figure \ref{fig:fiducial_tgas_evol}, but for our Low Mass model (see Table \ref{tab:Model_properties}). }}
    \label{fig:low_mass_tgas_evol}
\end{figure*}

\begin{figure}
    \centering
    \includegraphics[width=0.8\linewidth]{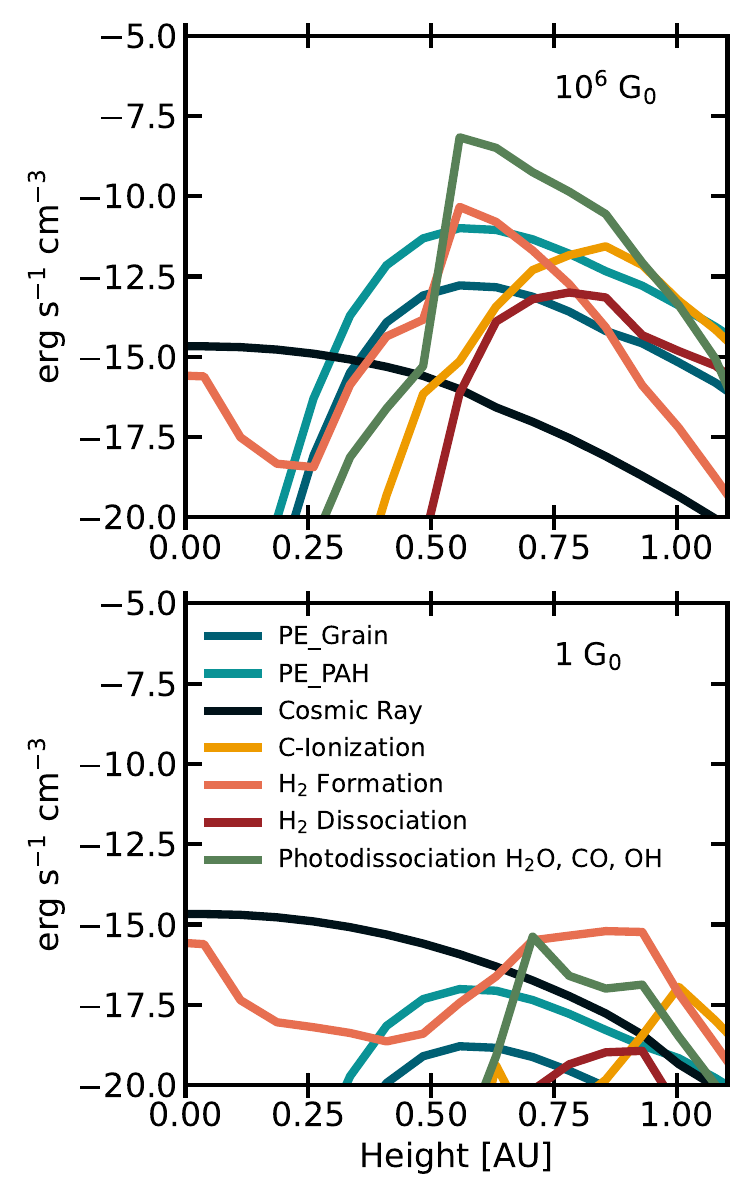}
    \caption{\textit{The primary heating mechanisms at 3~AU within a highly irradiated disk [top] and isolated disk [bottom]. PE stands for photoelectric effect. Most heating mechanisms within DALI are described in \citet{Bruderer12,Bruderer13} and references therein, and the photodissociation of water, CO, and OH was added recently, as described in \citet{Bosman22a}.}}
    \label{fig:heating_cooling}
\end{figure}

\subsection{Temperature Effects}\label{sec:temp}
We first explore the impact of the external radiation field on the temperature structure of the disk. As we increase the UV background intensity, the temperature beyond 1~AU and in the atmosphere increases most significantly (see Figures \ref{fig:fiducial_tgas_evol} and \ref{fig:low_mass_tgas_evol}). This increase in the atmosphere is due to the photodissociation of 
H$_{2}$O, CO, and OH, photoelectric heating via grains and PAHs, as well as heating from carbon ionization, and heating from H$_{2}$ pumping, formation, and dissociation (see Fig. \ref{fig:heating_cooling}). In the \midUV{} model, the increase in temperature as compared to an isolated disk reaches a factor of two above a z/r of 0.25 and beyond 3~AU. When exposed to a UV background of 10$^{6}$G$_{0}$, this same region now exhibits temperatures well above 500~K, representing up to a factor of 10 increase in temperature. Effects are seen towards the midplane as well, but only beyond $\sim$1~AU,  for example, at 4~AU the temperature increases from 25~K in an isolated disk to 50~K when exposed to an extreme UV background. 

The trends seen in the fiducial disk model are enhanced for our low-mass disk case, because the UV-enriched area now penetrates deeper into the disk. When exposed to a UV background of 10$^{4}$G$_{0}$, the temperature above a z/r of 0.2 increases by a factor of two or more, and even at this `moderate' UV exposure, there is a 20\% increase in temperature down to the midplane starting at $\approx$7~AU. In our low mass model, exposed to 10$^{6}$G$_{0}$, the temperature in the disk beyond 1~AU and above a z/r=0.1 more than doubles, with a factor of ten increase for any part of the disk beyond $\sim$5~AU, including the midplane. 



\begin{figure*}
    \centering
    \includegraphics[width=0.8\linewidth]{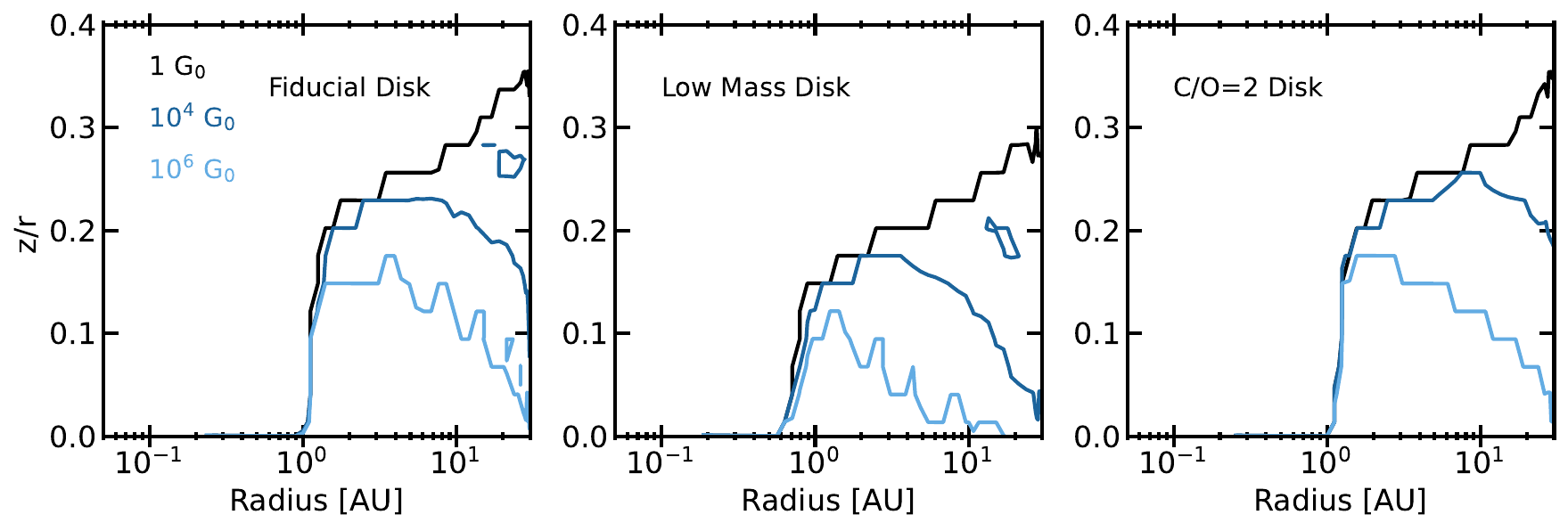}
    \caption{\textit{The different locations of the water snow surface across different UV irradiation environments. With increasing UV background, the snow surface is pushed deeper into the disk, and the secondary intersection with the midplane is moved inwards. }}
    \label{fig:snowlines}
\end{figure*}

Snow surfaces trace the location at which a molecule undergoes a phase change, corresponding to the temperature at which a given molecule will sublimate off of dust grains. The water snow surface is a critical feature of a planet-forming disk, as the location and shape of this snow surface dictates where the ice composition of grains will change significantly, thus will impact the fragmentation velocity of grains, and can cause a build-up of pebble material, kick-starting planet formation \citep[i.e.][]{Drazkowska17,Schoonenberg17,Hyodo19,Drkazkowska23}. The snow surface for any molecule will intersect with the midplane twice, once in a higher density environment, close to the star at the sublimation temperature of that molecule; and a second time in the diffuse outer extent of the disk, where the disk is externally heated and reaches that sublimation temperature once again. \citep[i.e.][]{Oberg15_DCOp}.


With increasing UV background, the disk beyond $\sim$1~AU increases in temperature. Thus, the snow surface of each species will turn over and intersect with the midplane at smaller radii and higher density than in an isolated disk. This leads to a 2D snow surface that is pushed deeper into the disk and the secondary intersection with the midplane moves radially inward as compared to an isolated or low UV background case. We show this feature with the water snow surface as an example in Figure \ref{fig:snowlines}. Here, we see that changes in the location of the water snowline are impacted by disk mass and UV background, but there is not a significant difference between the fiducial case and high C/O case. In all cases, the second intersection of the transition from gas-phase to ice-phase water occurs beyond 10~AU, even in our most extreme UV irradiation case. Thus,  there remains a water ice-rich `cometary' region between $\sim$1~AU to 10~AU or more, however this region decreases in size as the background UV irradiation increases. 

\subsection{Chemical Impact On Commonly Observed IR Molecules}

\begin{figure}
    \centering
    \includegraphics[width=\linewidth]{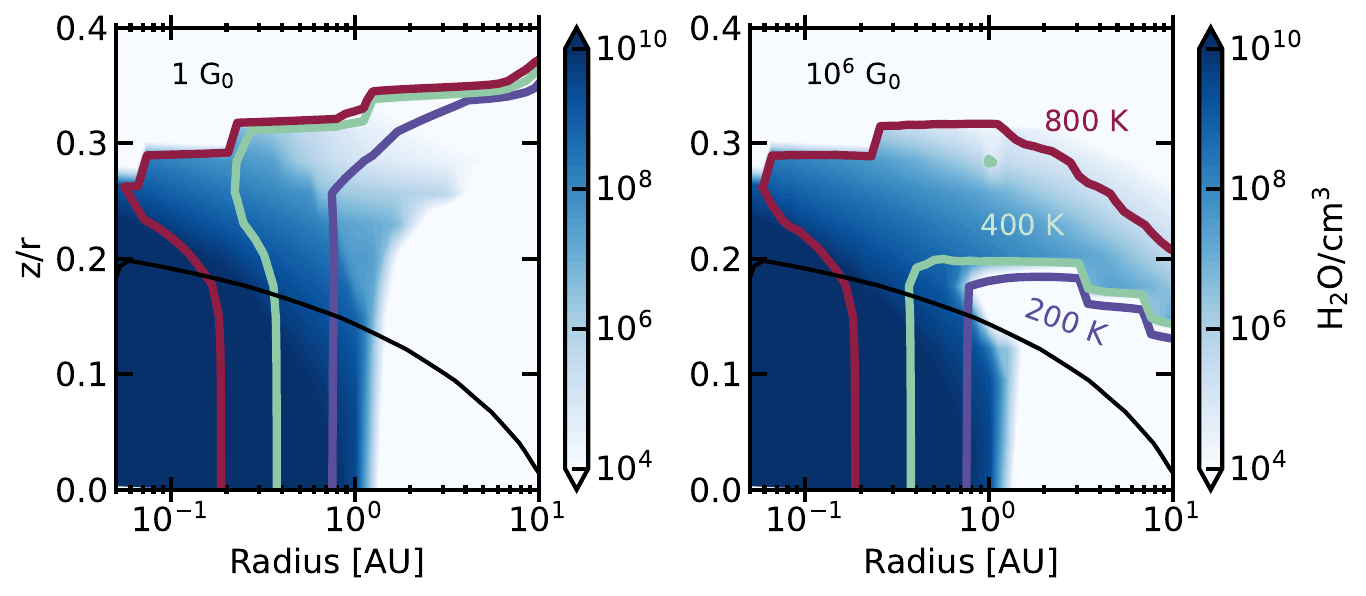}
    \caption{\textit{The 2D abundance of gas-phase water within an isolated disk case[left] and an irradiated disk[right]. The optically thick dust surface is marked in black contour, while isothermal contours are marked in colored lines (200, 400, 800~K).}}
    \label{fig:water_abundance}
\end{figure}


In addition to changing the temperature structure, external irradiation also directly impacts the disk molecular composition through increased photodissociation and photochemistry. Within this section we explore the impact of an irradiated environment on the disk chemistry that could be observable at IR wavelengths. 
With \textit{Spitzer} and now JWST, 100+ disks have been observed. The  most commonly identified molecules in protoplanetary disks are H$_{2}$O, OH, CO$_{2}$, C$_{2}$H$_{2}$, HCN, and CO \citep{Henning24, Arulanantham25}. These molecules have also been seen in  observations of UV irradiated disks \citep{RamireTannus25}, thus we focus on the potentially-observable changes of these molecules. In the isolated disk case, H$_{2}$O, OH, CO$_{2}$, and C$_{2}$H$_{2}$ all primarily exist within the gas-phase within 1~AU\citep{Arulanantham25}, where we do not see much of a change of temperature nor internal UV field. For all molecules, we see that the photodissoication front is pushed deeper into the disk as the UV background increases, and the location of the peak abundance in the gas-phase is moved downwards (see Fig. \ref{fig:water_abundance} for an example). The impact that the UV field has on each molecule is not only modulated by the physical structure (i.e. where the UV-optically thick surface is), but could also be influenced by the assumed C/O ratio throughout the disk.

\begin{figure*}
    \centering
    \includegraphics[width=0.9\linewidth]{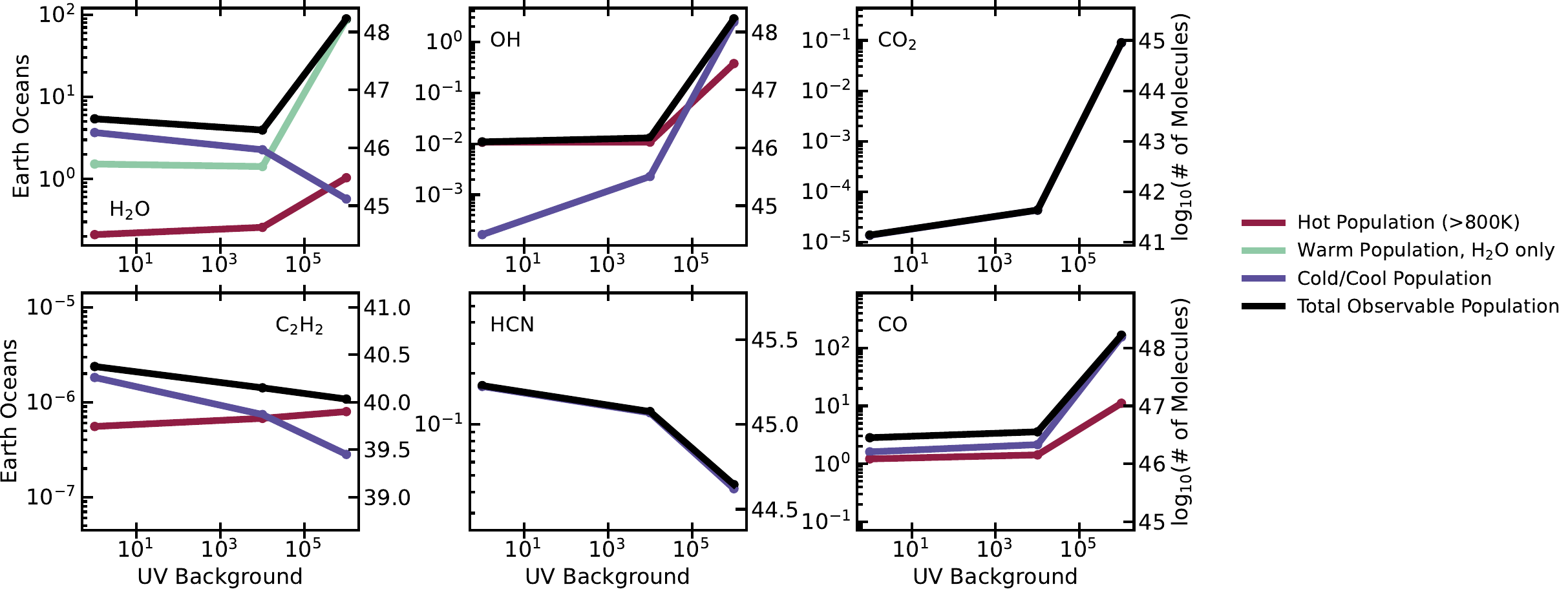}
    \caption{\textit{The evolution of key observable molecules with JWST over UV background in the fiducial model. One Earth's Ocean is equivalent to 10$^{46 }$ molecules, and is calculated using integrated column densities within a given temperature range, above the dust optically thick surface, and across all applicable radii. The cold population for water contains water molecules between 200-400K, while the cool population for all other molecules is 300-800~K. For CO and HCN the cool population dominates the total.}}
    \label{fig:fiducial_evol}
\end{figure*}

\begin{figure*}
    \centering
    \includegraphics[width=0.9\linewidth]{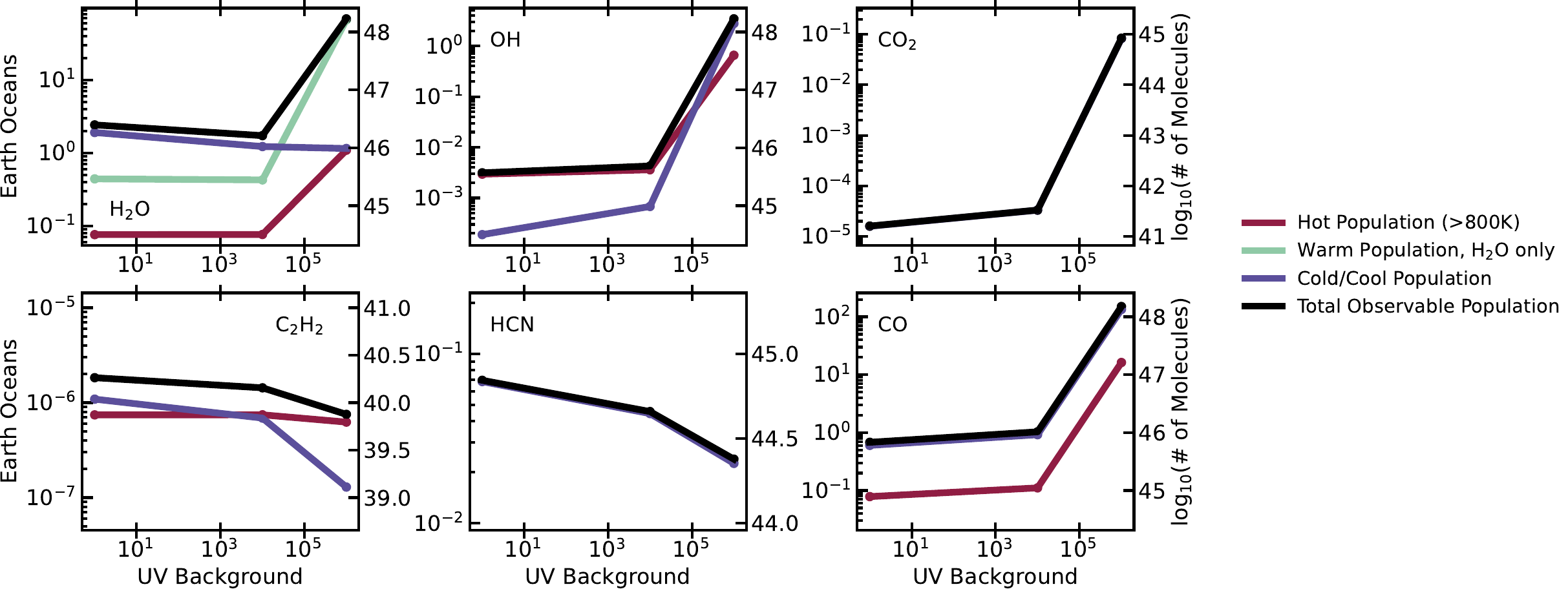}
    \caption{\textit{Same as Figure \ref{fig:fiducial_evol}, but for the low mass model.}}
    \label{fig:low_mass_evol}
\end{figure*}

\begin{figure*}
    \centering
    \includegraphics[width=0.9\linewidth]{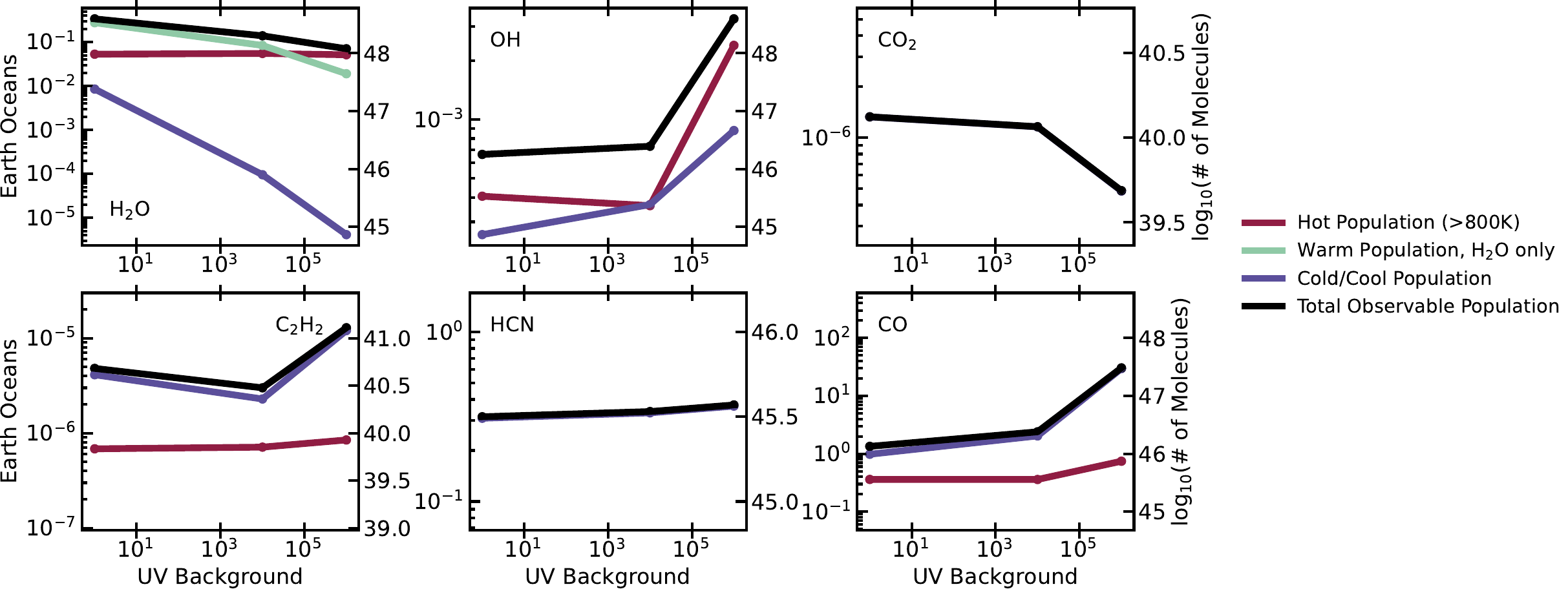}
    \caption{\textit{Same as Figure \ref{fig:fiducial_evol}, but for the C/O=2 model.}}
    \label{fig:c/o_evol}
\end{figure*}

We present our results as IR-emitting column densities in observationally motivated temperature bins. These are calculated by using the 2D abundance plots of each molecule, and determining the reservoir of each species identifying the `hot' population and the `cool' population. Motivated by slab model results and retrieved temperatures, we define a `hot' population as being above 800~K, and a `cool' population between 300-800~K \citep[i.e.][]{Arulanantham25}. Water has shown to exhibit a wider span of temperatures, thus for water we segregate into a hot, warm, and cold population \citep{RomeroMirza24,Banzatti25,Temmink25}. `Hot' water is anything above 800~K, `warm' is between 400-800~K, while `cold' is between 200-400~K. Fig. \ref{fig:water_abundance} shows where these boundaries lie within the disk with respect to the water abundance. We disregard the molecular abundance beyond an H$_{2}$ column density of 10$^{24.2}$ cm$^{-2}$, as this corresponds to where the disk becomes optically thick to 10$\mu$m emission in our models. The results for the commonly observed molecules are shown in Figure \ref{fig:fiducial_evol}, \ref{fig:low_mass_evol}, and \ref{fig:c/o_evol}. 

\subsubsection{Oxygen Carriers: Water, OH, and CO$_{2}$}

The oxygen-bearing molecules appear to be most strongly impacted by the UV background. H$_{2}$O, OH, and CO$_{2}$ increase in abundance when exposed to a high UV background if they naturally extend beyond 1~AU in the atmosphere in an isolated disk. Both the warmer disk temperature and the increased internal UV environment push the extent of the environment in which gas-phase water, OH, and CO$_{2}$ exist. In our oxygen-rich models (where C/O=0.47), the key gas-phase oxygen carriers all increase in potentially-observable abundance once exposed to a UV field greater than \midUV{}. At lower UV, the region of the disk $>$150~K does not extend beyond what is seen in an isolated disk. Isolating trends in each temperature component, the moderate temperature population of each molecule is increasing most significantly with higher UV field (see Fig. \ref{fig:fiducial_evol}, \ref{fig:low_mass_evol}). 

There is always a region in the disk where the UV-field is not strong enough to photo-dissociate \HtwoO, and not cool enough to freeze it out onto grains. This can been deemed as the `water-rich' zone. As the UV field increases, the combination of a higher UV field within the disk and higher temperatures in the atmosphere and outer disk increases the area of this `water-rich' zone, also increasing the total potentially-observable abundance. A similar molecule-rich zone is seen in the OH and CO$_{2}$ populations. The extent of the cold water (200-400~K) appears to decrease at \highUV{} as compared to an isolated disk. This is because temperature gradient from 200-400~K occurs over a smaller spatial scale in a highly irradiated disk. In 3D space, it will also be surrounded by warm water, likely blanketing the cold water population, hiding it from view due to water self-shielding \citep{Bosman22b}.  


In the carbon-rich model, we see a different trend  across all three O-containing molecules; the change over UV background is less extreme and leads to a slight decrease in the H$_{2}$O and CO$_{2}$ IR-emissive abundances (see Fig. \ref{fig:c/o_evol}). In order to achieve a C/O ratio of two, we depleted oxygen and kept the carbon abundance the same. Thus, our oxygen bearing species are less abundant than in our C/O = 0.47 case (see Table \ref{tab:Model_properties}). 
Now, with a lower total abundance of O-bearing molecules in the atmosphere of the disk changes in column density beyond 1~AU of especially H$_{2}$O and CO$_{2}$ are subtle. In a carbon-rich environment, what little water and CO$_{2}$ there is can be photodissociated, and some of their oxygen components will be distributed into CO or other C-bearing molecules.  However, due to such a low abundance, and low relative change in abundance from an isolated disk to a highly irradiated case, it is unlikely that any changes due to UV background will be detectable in a C/O=2 case. OH on the other hand, primarily exists within the warm atmosphere of the disk where it is still impacted by the UV field, and is a byproduct of photodissociated water. The OH-rich region is pushed deeper into the disk as it is exposed to higher UV fields. At \highUV{} the IR-emissive OH abundance increases by an order of magnitude, and according to these models, may be the only molecule that traces the increased UV field.  

\subsubsection{Carbon Carriers: C$_{2}$H$_{2}$, HCN and CO}

For this work, we consider C$_{2}$H$_{2}$ and HCN as our representative carbon carrier molecules as these are commonly seen with JWST and are often used as a tracer of the carbon content of the disk \citep{Tabone23,Colmenares24}. In all models, the bulk of the gas-phase \CtwoHtwo{} exists in the warm atmosphere, primarily within 0.1~AU which is very well shielded from the external UV field. In our models where C/O=0.47, there is a very small abundance of IR-emissive \CtwoHtwo{}, and as the UV field increases, its total abundance decreases by a factor of $\sim$two. We do not expect these subtle changes to be detectable with JWST. In the high C/O case, there is more \CtwoHtwo{} in the atmosphere, which can be impacted by the UV background, and there is an increase in IR-emissive \CtwoHtwo{} in the most extreme UV case. Overall, however, these changes continue to appear relatively small, only by a factor of up to three and it is unclear if these will be distinguishable amongst other sources of uncertainty in these sources i.e. variability, natural chemical diversity, or just the sensitivity of JWST. 

\begin{figure}
    \centering
    \includegraphics[width=0.8\linewidth]{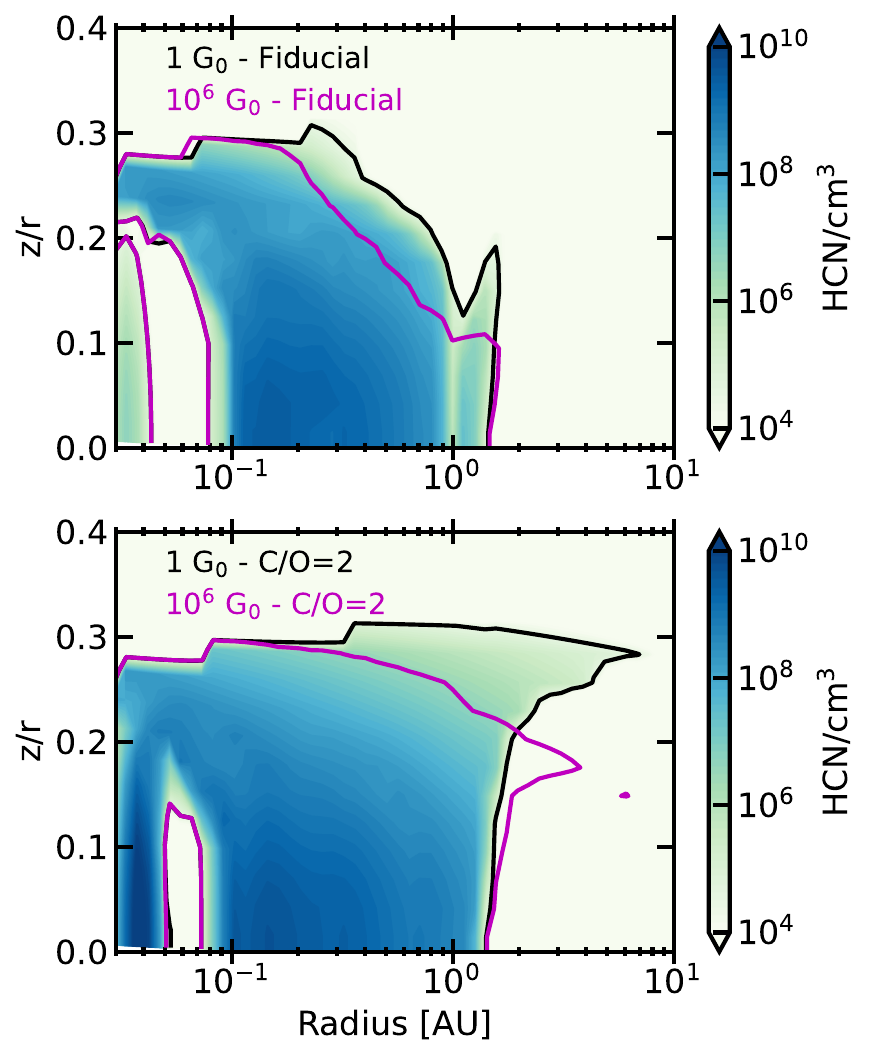}
    \caption{\textit{The abundance of HCN in our fiducial (top) and C/O=2 (bottom) models. The black contour traces an abundance of 1$\times$10$^{4}$ HCN/cm$^{3}$, inside of this contour contains $\approx$99.99\% of the HCN total abundance. The magenta contour traces the same HCN abundance, but in the \highUV{} models.}}
    \label{fig:HCN}
\end{figure}

HCN is one of the most abundant observable molecules within a disk, and is detectable both within the radio and IR regimes \citep{Bergner19, Arulanantham25}. It is highly abundant from the atmosphere down to the midplane within 2~AU, and extends out to $\sim$10~AU in the atmosphere of the C/O=2 model (see Fig. \ref{fig:HCN}). Similar to water, HCN is photodissociated deeper and deeper within the disk with increasing UV background. 
IR-emissive HCN abundance decreases with UV field (Fig. \ref{fig:fiducial_evol}, \ref{fig:low_mass_evol}), because HCN is destroyed by the increased level of O and OH in the atmosphere of an irradiated disk, which can explain the excess CO and CN in the space in which HCN used to occupy. In the high C/O case (Fig. \ref{fig:c/o_evol}), there is a near constant amount of potentially-observable HCN. This is because while some HCN is destroyed through the increase in OH in the atmosphere, there is more HCN in the outer disk, which is now potentially-observable with JWST after being exposed to a \highUV{}. Throughout all irradiation backgrounds, the cool HCN (300-800~K) makes up the vast majority of what is IR-emissive. While this cool population of HCN is spread out, occupying a larger spatial extend over higher UV background, what is IR-emissive remains at a nearly constant abundance. 

CO is highly abundant in the gas-phase of the disk, and its IR-emissive abundance increases by at least a factor of 10 in each model. This increase in potentially-observable abundance is primarily due to the increase in the spatial extent of the disk that is between 300-800~K. CO flux is likely impacted by the UV background, thus using CO as a comparative molecule in chemical contexts must account for the associated irradiation environment. 

\subsubsection{Column Density Ratios}

\begin{figure*}
    \centering
    \includegraphics[width=0.8\linewidth]{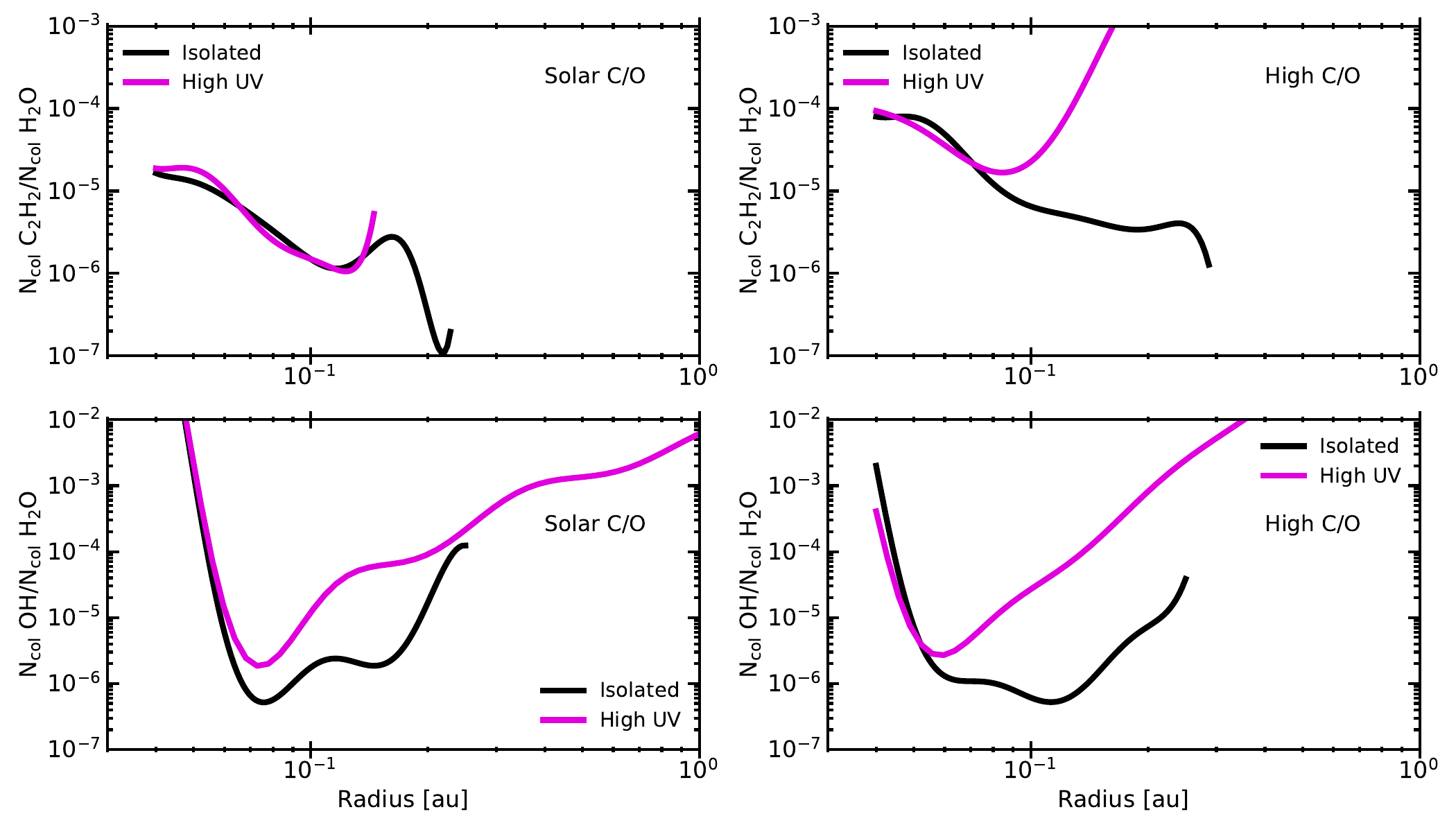}
    \caption{\textit{The ratio of the potentially-observable column density of \CtwoHtwo{} and \HtwoO{} [top] and OH over \HtwoO{} [bottom]. The observed ratios can be altered not only by initial C/O ratio of the gas, but also irradiation background (Isolated disk represented by a black curve, and a \highUV{} by magenta).}}
    \label{fig:Ncols}
\end{figure*}

The ratio between the derived column densities of \CtwoHtwo{} and \HtwoO{} are often used to probe the C/O ratio in the inner disk \citep[i.e.][]{Tabone23,Colmenares24}. We can use our models to derive an observed column density of these molecules and see how these observable quantities change with UV background. In Figure \ref{fig:Ncols}, we see that, as expected, with an increase of the C/O ratio, the column density ratio of \CtwoHtwo{}/\HtwoO{} increases in both an isolated disk and highly irradiated case. In an irradiated disk, the ratio increases even more significantly. However, in the case of a solar C/O ratio, the observed column density ratio of C$_{2}$H$_{2}$ and \HtwoO{} does not change over increased UV background. This implies that an observed high C/O ratio can only be due to a carbon-rich environment, and a highly irradiated background can enhance that observed C/O value. In a highly irradiated disk, it may be difficult to determine an exact C/O ratio without alternative constraints on the UV field, given this combination of the C/O $>$ 1 and high UV field both increasing the ratio between the column densities. 

Another ratio we expect to change with UV field is OH/\HtwoO{} \citep{Tabone23}. Under high UV radiation, regardless of the C/O value, the OH abundance beyond 0.1 AU increases, so that the OH/\HtwoO{} ratio increases as well. In a \highUV{} the OH and H$_{2}$O column likely can be measured beyond 1~AU, as opposed to an isolated disk where it is constrained to well within 1~AU. With increasing UV field, \HtwoO{} will be photodissociated, creating an enhancement of OH which will further break down in to atomic oxygen. In both the solar C/O model and C/O = 2 model, the O/H ratio changes over UV background. However, the column density ratio of OH/\HtwoO{} is not altered between the C-poor and C-rich models; nearly identical trends are seen when a highly irradiated background is introduced.  

\subsection{Chemical Changes Underneath the Surface}

\begin{figure*}
    \centering
    \includegraphics[width=0.8\linewidth]{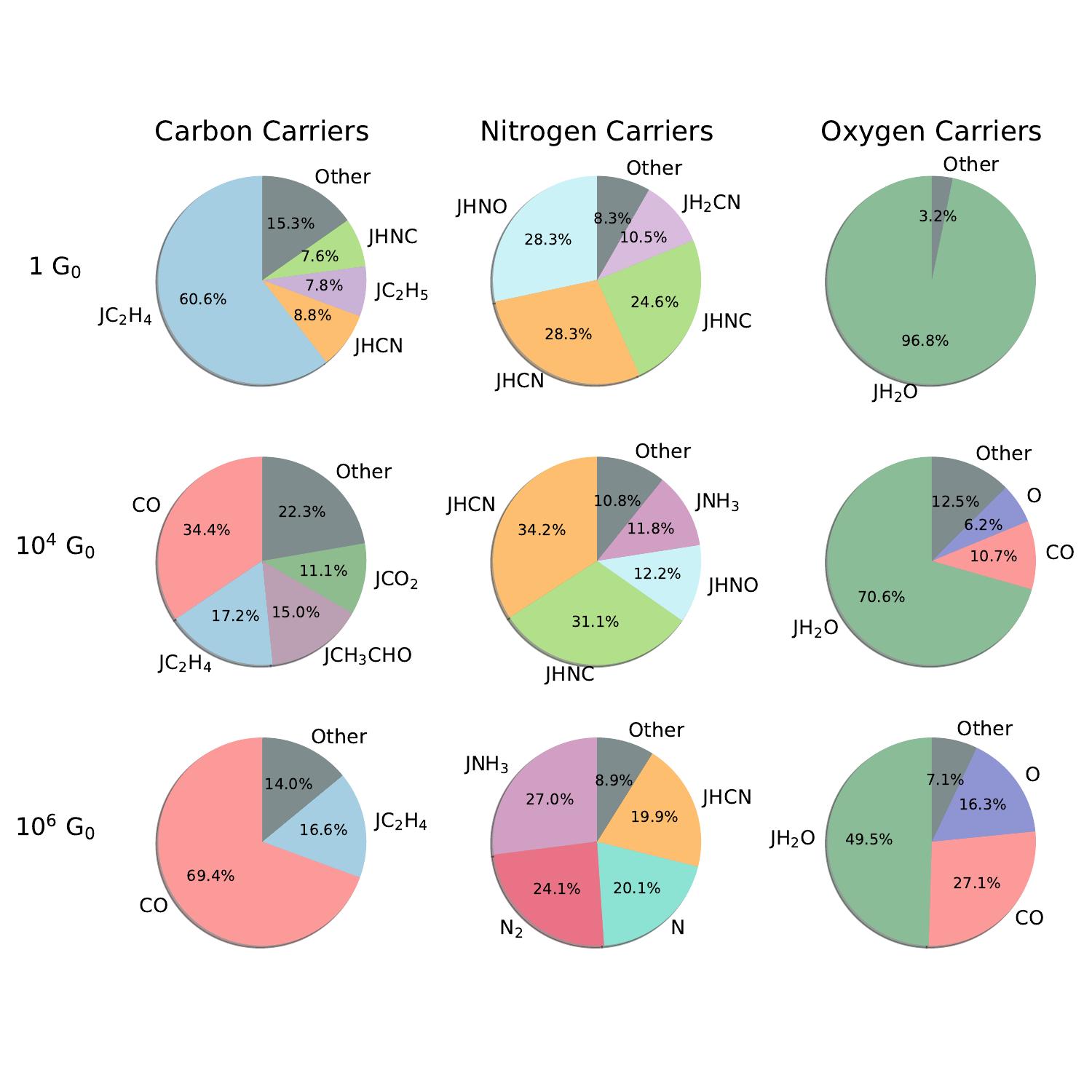}
    \caption{\textit{The primary carbon, nitrogen, and oxygen carriers near the midplane between 1-10~AU in our fiducial model and how they evolve over different UV backgrounds. A molecule starting with ``J'' denotes it is in the ice-phase.}}
    \label{fig:pie_charts}
\end{figure*}

In the previous sections, we were motivated by the impact of an increased UV-field within the observable region of the protoplanetary disk. Now we focus on chemical impacts `underneath the surface'. JWST is sensitive to the atmosphere of the disk (in our models z/r>0.3). Not until a \highUV{} does the disk at z/r=0.2 get warm enough to be emissive in the IR. This region is just the tip of the iceberg of most molecular reservoirs. There are chemical changes closer to the planet-forming midplane that come from an increase in the UV-background below the region of the disk that is optically thick at IR wavelengths. Figure \ref{fig:pie_charts} shows the evolution of the main carbon, nitrogen and oxygen carriers near the midplane between 1-10~AU. 

Since inward of 1~AU there is no change in the temperature nor UV background, we focus on changes in the O, C and N carriers near the midplane (z/r<0.15) between 1-10~AU which is where we expect most gas-rich planets to form, and which coincides with the cometary region of our Solar System. Starting with an isolated disk, frozen-out water is the primary oxygen carrier, carrying over 95\% of the total oxygen. With increasing UV background, the temperature increases, sublimating more and more water so that gas-phase CO becomes a more dominate O-carrier in the midplane. Due to the increase in the UV, atomic oxygen also increases in abundance and has the largest relative increase compared to other oxygen carriers. Amongst the nitrogen-bearing species, in our isolated disk, most N is frozen out onto grains and is primarily partitioned into HCN, HNO, and HNC, although this may be dependent on the breadth of the nitrogen carriers in our chemical network, and that we initialize our chemistry to be in the atomic phase. With increased UV these ice species remain the primary N carriers, but atomic gas-phase N takes up a larger and larger fraction of the total N reservoir. In the most extreme UV irradiation case, nearly half of the nitrogen is now in the gas-phase in either N$_{2}$ or N, and much of the ice phase is in NH$_{3}$. Amongst the carbon-carriers, much of the carbon that is not in CO is in a variety of ice-phase hydrocarbons which act as a carbon sink (C$_{2}$H$_{4}$). The specific molecule that acts as a carbon sink will change based on the breadth of the chemical network, but the existence of a carbon sink will be present regardless of the chemical network. For example, when C$_{2}$H$_{2}$ is set as the largest hydrocarbon in the network, there is an appreciable abundance of it in the midplane near 1~AU, with an expanded network, carbon is redistributed to larger hydrocarbons. This likely is a feature of the hydrocarbon chemistry network in our model and the fact that we initialize our initial chemistry as atomic species. Future work is required to confirm whether this is a realistic scenario. 

Near the midplane, the chemical changes due to the increased UV background are brought about both by an increased temperature and UV field. More of the carbon, nitrogen, and oxygen carriers become a part of the gas-phase, and those that are in the gas-phase break down into more simple parts such as atomic N and O. Thus, what is feeding forming planets between 1-10~AU in a highly irradiated disk will be overall ice depleted, still water rich, and with a more atomic-rich gas than compared to an isolated disk. 


\section{Discussion}\label{sec:diss}


There are changes to the chemistry of planet formation in an irradiated disk as compared to an isolated. These changes take place in the disk atmosphere and primarily beyond 1~AU.
With common assumptions of disk physical structure (gas-to-dust ratio, stellar-to-disk mass ratio, small-to-large grain dust distribution), supported by observations, the inner AU is well-shielded from even a \highUV{}. However, there is a significant chemical impact just beyond 1~AU, which will influence the inner 1~AU once dynamics are taken into account (i.e. radial drift of material) and the chemical composition of the atmosphere will be impacted strongly enough to detect beyond \midUV{}. Alterations of the chemistry present within the atmosphere also can influence the final chemistry of forming planets, through dynamical processes such as meridional flows onto actively forming gas giant planets. We continue to explore the impact from the UV background on the chemistry of planet formation, and compare our results to current observations in the following section.

\subsection{Snow Surfaces in Irradiated Disks}

The water snow surface is an influential boundary within protoplanetary disks. The bulk of the ice mantel surrounding dust grains are made of water, thus once water can sublimate off into the gas phase, the overall material property of the grains change, and entrapped molecules such as CO, CH$_{3}$OH, and other COMs can sublimate as well and enhance the inner disk gas reservoir \citep[i.e.][]{Collings04,Simon23}. After being exposed to an enhanced UV background, the inner boundary of the water snow surface does not change, remaining near 1~AU for each model. Within the water-ice rich region of our models, water remains the primary oxygen carrier, but hydrocarbons sublimate off of grains, and add to the gas-phase CO population as the UV background increases.  
While the inner boundary of the water snow surface does not change in our models over different UV backgrounds, the outer boundary, where it turns over and hits the midplane, does change with UV background, getting as far in as 10~AU in our low mass model. The grains beyond the outer boundary of the water snow surface will be relatively bare, and hence with increasing UV background, a larger percentage of the grain population will be bare. These bare grains are more prone to fragmentation\citep{Blum08, Gundlach11,Gundlach15, Musiolik16}. Not only will external radiation impact the ratio of ice/gas for life-critical molecules, but stripping grains of water ice mantels will impact dust dynamics and growth.  

The fact that the water snow surface also decreases in scale height with increasing UV background has implications for the `cold finger' effect. The cold finger effect starts with the process of gaseous water just above the snow surface diffusing into the freeze-out region. Once frozen-out onto grains, that water is depleted from the upper extent of the disk, and is recycled into the deeper regions of the disk as dust drifts settles downwards and drifts inward and is released into the inner disk. Given that the spatial extent of the water freeze-out region is smaller, this cold finger effect could be lessened dramatically, and what has been assumed as an effectively vertical snow surface (the atmospheric snowline and midplane snowline are at the same radial location) may not exist in highly irradiated disks \citep{Bosman21}. This means that there is no sink term for gas-phase water above the snow surface, and there is a less significant enhancement of water ice near the midplane. 

Our model is truncated at 30~AU, as we are motivated to understand the chemical changes to planet formation conditions within the inner few AU. Since the DALI's radiative transfer of the UV background only takes into consideration the column densities in the z-direction, we can truncate the disk in this way, and have the same results as if the disk extended to 100~AU or more. This is a reasonable simplification of the radiative transfer of the UV background impinging on the disk, as what is falling perpendicular to the disk will be most impactful, as it accounts for the strongest intensity interacting with the smallest optical depth. Any radiation coming from outside of the disk, parallel to the midplane, has to penetrate the the outer disk physical structure, and thus will have a negligible effect on the radiation field and resultant temperature and chemistry within the inner disk. Thus our results should be impacted by our truncated model.

\subsection{Reset Chemistry at the Midplane in Highly Irradiated Disks}

\begin{figure*}
    \centering
    \includegraphics[width=0.9\linewidth]{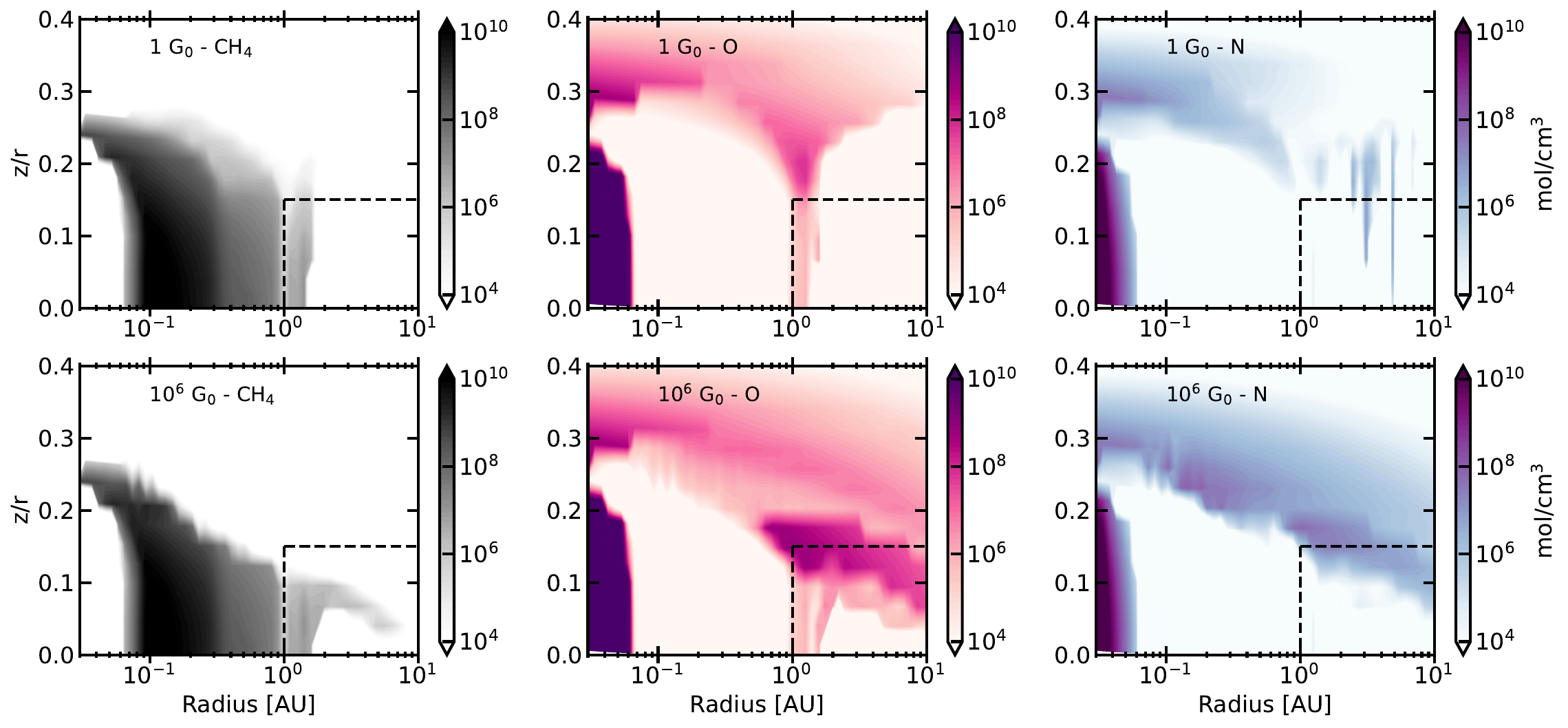}
    \caption{\textit{The abundances of CH$_{4}$, atomic O, and atomic N in an isolated disk [top], and disk exposed to \highUV{}. Changes in abundance close to the midplane are particularly striking in some of the most simple C, N, and O carriers. The dashed line marks the portion of the disk that is accounted for in Figure \ref{fig:pie_charts} }}
    \label{fig:extra_molecules}
\end{figure*}

Beyond $\approx$1~AU in each model, the UV background penetrates down to the midplane, and breaks down molecules into simpler constituents and heats up the midplane marginally enough to sublimate some molecules off of grains (see Fig. \ref{fig:pie_charts}). Atomic nitrogen and oxygen become $>$15\% of the N and O budgets at the midplane in the most irradiated case.  
The increase in oxygen and nitrogen atoms is prevalent in the atmosphere and is halted at around a z/r=0.1 beyond 1~AU, and an excess of CH$_{4}$ resides just below this layer which represents one of the largest relative increases in a molecular abundance once a disk is exposed to a high UV (see Figure \ref{fig:extra_molecules}). CH$_{4}$ can form in the gas-phase in warm environments where the gas is oxygen-poor, where there is atomic carbon and CO \citep{Bergin95,ChaparroMolano12}. Otherwise, it forms efficiently via hydrogenation while frozen-out onto grains\citep{Oberg08}, and both pathways are accounted for in our chemical network \citep{Woodall07}. However, in our model the gas-phase pathway is what accounts for the excess CH$_{4}$ near the midplane beyond 1~AU. These factors for gas-phase formation  of CH$_{4}$ (warm temperatures, a source of atomic C, CO) exists within the inner AU in all models. In a highly irradiated disk, these factors are also present near the midplane between 1 and 10 AU.  The higher abundances of O, N, and CH$_{4}$ just above the midplane between 1-10~AU are due to the photodissociation front being pushed deeper into the disk. This can be seen as a reset chemistry, breaking down any chemical evolution in the outer disk/cloud that might have been inherited from a natal cloud and envelope if infall of material into this cometary region (1-10~AU) is not fast enough to replenish molecules being photodissociated. Similar results are shown in models of the outer disk chemistry exploring up to 1000 G$_{0}$ \citep{Gross25}, where CS and C$_{2}$H decrease in column density beyond 50~AU while C abundance increases. The atomic carbon abundances in our models also increase with UV background, however that is only seen in the low density atmosphere. 
Gaseous CO and N$_{2}$ also increase in relative abundance near the midplane in the high UV cases, however this result is less chemical in origin, and rather thermal and just a movement of their snowlines. These molecules have very low freeze out points, thus in highly irradiated disks effectivley do not have a snow surface.

\subsection{Winds, Photoevaporative Envelopes, and Dynamics}


In our modeling framework, we do not include photoevaporative winds, external shock fronts, nor envelopes.  Recent work has explored the impact of a photoevaporative wind surrounding a disk, exposed to up to a 5000 G$_{0}$ \citep{Keyte25}. They find that the presence of a wind shields the external UV field slightly, impacting the UV flux even down to the midplane. They find that these winds will impact observations, increasing the flux in the FIR of CO, OH, and atomic lines due to disk heating via thermal re-emission from the disk wind. Water fluxes appear to be negligibly impacted, but the abundances increase slightly in the presence of a wind. This work was focused on outer disk chemistry, thus it is difficult to say how these results will transfer into the inner disk, and likely strongly depend on the physical assumptions of the wind. 

We can approximate what material in our disk will be blown away in a wind based on the temperature and density. In disks exposed to a high UV background, the outer disk becomes warm, and can approach temperatures of 10$^{4}$~K. Near this temperature, the associated sound speed becomes comparable to the orbital velocity within the disk. When the sound speed of this hot gas is greater than the orbital velocity, the disk will photo-evaporate \citep{Alexander14, Winter22,Garate24}, and can impact the observed trends that this model predicts. However, we find even in our high UV case, the gravitational radius is greater than 10~AU, thus our model results focused on the inner disk perspective should not be impacted by photoevaporative winds due to high temperatures.

A critical limitation of this study is that the thermo-chemical model is static. There are known impacts due to dust and gas dynamics that should be taken into account simultaneously in order to have a well-informed theory of how the UV background will impact the chemistry of planet formation. In a \highUV{} low mass disk, beyond 10~AU the disk is warm enough to have water primarily in the gas phase. This leaves ice-free dust grains in the outer disk, and it has been shown that without an ice mantel, these bare grains may be more prone to fragmentation \citep{Blum08, Gundlach11,Gundlach15, Musiolik16}. In isolated systems with no perturbations in gas density, it is assumed that water ice is effectively delivered from the outer disk, into the inner disk where it sublimates. In this scenario, where there is only a small spatial extend of the disk that contains water ice, this delivery may not be so efficient. But future modeling work combining thermo-chemistry and dust and ice evolution is necessary to understand the extent of this effect. Beyond dust evolution, the presence of turbulence in the disk can impart chemical signatures from the atmosphere to the planet-forming midplane. As we increase the UV background, a higher percentage of the disk atmosphere becomes ionized. This will enhance the reach of the magneto-rotational instability or MRI and subsequently increase turbulence \citep{Okuzumi11}. Observational confirmation of this could come from the outer disk, as these effects will first be seen on larger spatial scales. Turbulence is a difficult value to constrain with observations, but there are promising avenues such as observing the emission surfaces of select molecules \citep{PanequeCarreno24}.

\subsection{Comparison to Current Observations}

With JWST there are now 13 disks that are exposed to high UV environment due to surrounding massive stars with published MIRI spectra \citep{RamireTannus25, Zannese25}. The sources within the XUE survey \citep{RamireTannus25}, are all Herbig or intermediate T Tauri disks (1-4 M$_{\odot}$), but exhibit spectra that are akin to typical T Tauri spectra. They contain water, CO, HCN, CO$_{2}$, and C$_{2}$H$_{2}$. The proplyd in Orion (d203-506) retains a similar spectra, but also detections of CH$^{+}$ and CH$_{3}$$^{+}$ \citep{Zannese25}, while those in the XUE survey do not. Based on these observations, it appears that JWST spectra do not vary significantly between isolated disks and the irradiated ones that have been sampled so far. 

We find a promising agreement between published irradiated disk spectra and our findings.We make comparisons of previous observations to results we present in Section \ref{sec:results}, which presents how the IR-emissive or potentially-observable population of each molecular species changes over UV background. It is worth noting that any changes in potentially-observable abundance are an upper limit on what could be observed, as molecules can become optically thick in the atmosphere of the disk, and thus the only detectable changes are above a given molecule's optically thick surface. Our models predict that any disk exposed to a less than \midUV{}, the potentially-observable abundances and temperatures do not change appreciably. Approximately half of the XUE sample appears to be exposed to \midUV{} or below, thus based on our models we would not expect any change in JWST spectra between an isolated and moderately irradiated disk. The XUE sample found no cold water in any of the disks, which our models can explain, as cold water will occupy a smaller fraction of the total water population and it is blanketed by warm and hot water. The proplyd disk d203-506 contained a signature of CH$_{3}^{+}$ while those in the XUE sample did not, suggesting this emission likely originated from the envelope. Our models only produce CH$_{3}^{+}$ within 1~AU, and thus we would not expect CH$_{3}^{+}$ flux to increase in irradated disks, like what was found in XUE. 

We do expect a significant relative increase in flux for OH, regardless of initial C/O ratio, which is found in the Orion proplyd, but there are only tentative detections of OH in the XUE sample. At  >\midUV{}, we would expect warm water lines to increase in intensity due to the significant increase in abundance, yet that is not found in any of the irradiated disks observed to date. Critically, we set out to model the typical chemistry within a disk around a 0.6 M$_{\odot}$ star, while the XUE sample and d203-506 are more massive \citep{Zannese25, Schroetter25}. We also lack a time component in our models, as these disks likely are not exposed to a \highUV{} for their entire lifetime. These, and the lack of modeling winds and evelopes can start to explain the differences between our model findings and observations. 
However, it is promising that some of our predicted water trends, and relatively unchanging abundances of C$_{2}$H$_{2}$ and HCN, seem to hold true in the observed externally irradiated sources. With only 13 sources at a wide ranges of projected distances from an irradiation source, we are still lacking the statistics to say what features of a disk's spectra is derived from the UV background as opposed to internal properties of each individual system.

\section{Conclusion}\label{sec:conclusion}

The inner few AU of a protoplanetary disk is a critical region for planet formation. It hosts the highest density dust and gas within the disk, the bulk of the terrestrial planets likely form within this region, and most known exoplanets reside within a few AU of their host star. Most planet-forming disks reside within highly irradiated environments. Thus, to understand the chemistry of `typical' planet formation, we must look towards the inner AU of disks that are within an enhanced UV environment. We have explored how the inner disk chemistry changes due to an irradiated background (1 G$_{0}$ - \highUV{}). We find that the bulk of the potentially-observable changes from the inner disk occur beyond \midUV{}, and such a UV field will significantly impact both the molecular reservoir accessible with the JWST, and the chemistry available to actively forming planets. Our main findings are:

\begin{itemize}
    \item As the UV field increases, the temperature structure of the disk increases. The outer disk becomes warmer, in the most extreme case, temperatures of $>$500~K are seen at moderate disk heights (z/r<0.2) out to 10~AU.

    \item Icelines of various molecules will `turn-over' in high UV environments, insinuating that the outer disk will be more gas-rich in irradaited disks. In the case of water, the ice-rich zone is constrained to between 1 to a few 10s of AU in highly irradiated disks. The surface itself is pushed down deeper into the disk. This may dampen the `cold finger' effect and impact dust evolution and growth.

    \item In the IR, water and OH lines are most significantly impacted by changing UV background. Cold water lines nearly vanish, while hot and warm reservoirs appear enhanced.

    \item Near the midplane, sublimation and short gas-phase reaction timescales result in a near reset of the chemistry. 

\end{itemize}

Larger surveys of irradiated sources, over different stellar types, are necessary to have direct evidence of the unique impact that the external irradiation will have on the chemistry of planet formation. Equally as essential are robust models that explore a wide parameter space, including the complexities of the internal systems (winds, envelope, truncated disks) as well as timescales and dynamics. This work demonstrates that highly irradiated environments are not detrimental to planet formation, but will alter the chemistry being inherited to planets forming in irradiated disks.


\section{Acknowledgments}

software: Matplotlib \citep{Matplotlib}, Astropy \citep{astropy:2013,astropy:2018}, NumPy \citep{harris2020array}. 

J.K.C. acknowledges support from the Kavli-Laukien Origins of Life Fellowship at Harvard. K.I.\"{O}. acknowledges  Simons Foundation grant No. 686302, and an award from the Simons Foundation (grant No. 321183FY19).  A.S.B. is supported by a Clay Postdoctoral Fellowship from the Smithsonian Astrophysical Observatory. The authors would like to thank the anonymous referee for their insightful comments.


\appendix
\restartappendixnumbering
\renewcommand{\thetable}{A\arabic{table}}

\section{DALI Physical Structure}

We model the physical structure of our disk by using a surface density that has been widely used in modeling protoplanetary disks\citep{Lynden-Bell74}: 

\begin{equation}
\Sigma(r)=\Sigma_{c}\left(\frac{r}{r_{c}}\right)^{-\gamma}\exp{\left[-\left(\frac{r}{r_{c}}\right)^{2-\gamma}\right]}
\label{surfden}
\end{equation}

\noindent where \(r_{c}\) is the characteristic radius at which the surface density is \(\Sigma_{c}\) and \(\gamma\) is the power-law index that describes the radial behavior of the surface density. The 2D density distribution is as follows: 

\begin{equation}
    \rho(r,z) = \frac{\Sigma(r)}{\sqrt{2\pi}h(r)} \exp{\left[-\frac{1}{2}\left(\frac{z}{h(r)}\right)^{2}\right]}
\end{equation}

\begin{equation}    
    h=h_{c}\left( \frac{r}{r_{c}}\right)^{\Psi}
\end{equation}

\noindent Where \(h_{c}\) is the scale height at the characteristic radius, and \(\Psi\) is a power-law index that characterizes the flaring of the disk structure. The gas and dust populations follow equations A1-A3, with the values in Table \ref{tab:struct_properties}.  Each dust population follows an MRN grain size distribution \(n(a) \propto a^{-3.5}\) \citep{Mathis77}.

\begin{deluxetable}{cc}
\label{tab:struct_properties}
\tablecolumns{7}
\tablewidth{0pt}
\tabletypesize{\small}
\tablecaption{Model Variables}
\tablehead{Factor	& Value }
 \startdata
 M$_{\rm{gas}}$	&	0.06, 0.06 M$_{\odot}$	\\
M$_{\rm{dust}}$/M$_{\rm{gas}}$	&	1000	\\
M$_{\rm{small dust}}$	&	M$_{\rm{dust}}$*0.01	\\
M$_{\rm{large dust}}$	&	M$_{\rm{dust}}$*0.99	\\
r$_{c}$	&	100 AU	\\
h$_{c}$ [gas and small dust]	&	10 AU	\\
h$_{c}$ [large grains]	&	1 AU	\\
$\gamma$	&	1.0	\\
$\psi$	&	1.05	\\
 \enddata
\end{deluxetable}

\bibliography{references}{}
\bibliographystyle{aasjournal}

\newpage

\end{document}